\documentclass[12pt]{article}
\usepackage[dvips]{graphicx}
\usepackage{wrapfig}
\usepackage{color}
\usepackage{boxedminipage}
\usepackage{amsfonts}
\usepackage{amssymb,amsmath}
\usepackage{url}
\usepackage{helvet} 
\usepackage{cite}

\makeatletter 
\renewcommand{\section}{\@startsection{section}{1}{0mm}{2mm}{2mm}{\Large\normalfont\bfseries}}
\renewcommand{\subsection}{\@startsection{subsection}{2}{0mm}{1mm}{1mm}{\large\normalfont\bfseries}}
\renewcommand{\subsubsection}{\@startsection{subsubsection}{3}{0mm}{1mm}{0.5mm}{\normalfont\bfseries}}
\makeatother 

\newcommand{\chivec}{\boldsymbol{\chi}}

\newcommand{\dg}{\Delta G}

\newcommand{\dgfold}{\dg_{\mathrm{fold}}}

\newcommand{\eq}[1]{Eq.\ (\ref{#1})}

\newcommand{\lamvec}{\boldsymbol{\lambda}}
\newcommand{\lav}{\left \langle}
\newcommand{\lb}{\left [}
\newcommand{\lp}{\left (}

\newcommand{\neff}{N^{\mathrm{eff}}}

\newcommand{\pmf}{\mathrm{PMF}}

\newcommand{\pfold}{p(\mathrm{folded})}

\newcommand{\punf}{p(\mathrm{unfolded})}

\newcommand{\rav}{\right \rangle}
\newcommand{\rb}{\right ]}

\newcommand{\rp}{\right )}

\newcommand{\tcstar}{t^*_{\mathrm{corr}}}
\newcommand{\tinit}{t_{\mathrm{init}}}

\newcommand{\tsim}{t_{\mathrm{sim}}}

\newcommand{\ulam}{U_{\lamvec}}

\newcommand{\xbf}{\mathbf{x}}

\newcommand{\aisgen}{Neal-2001,Huber1997}
\newcommand{\aisdmz}{Lyman2007,Lyman2009}

\newcommand{\assessbrief}{Mountain1989,Straub1993,Neirotti2000,vanGunsteren-2002,Hess2002,Lyman2007a,Grossfield2007a,Grossfield2009,Zhang2010}

\newcommand{\dmzbook}{Zuckerman2010}

\newcommand{\GrossfieldZuckerman}{Grossfield2009}

\newcommand{\lambdadyn}{Tidor-1993,Brooks-1996}

\newcommand{\mcdyn}{Nowak2000,Shimada2001,Zuckerman-2004d}
\newcommand{\multicanon}{Hansmann1996,Mitsutake2001}

\newcommand{\pmfrefs}{Swendsen-2004,Shirts2008,Maragliano2008,Abrams2008,Zheng2008a}

\newcommand{\repexalgo}{Swendsen-1986,Geyer1991,Nemoto-1996}
\newcommand{\repexcrit}{Zuckerman2006,Denschlag2008,Nymeyer2008}
\newcommand{\repexdistrib}{Rhee2003}
\newcommand{\repexopt}{Kofke2002,Rathore2005,Trebst2006,Sindhikara2008}
\newcommand{\repexvar}{Opps2001,Rhee2003,Fenwick2003,Garcia-2004,Berne-2005,Calvo2005,Li2006b,Rick2007,Okur2007,Brenner2007,Ruscio2010}
\newcommand{\resexdmz}{Zuckerman-2006a,Zuckerman-2006c,Zuckerman2009}
\newcommand{\resexvar}{Christen2006,Liu2008b}

\begin{document}

\noindent
Title: Equilibrium Sampling in Biomolecular Simulation\\
(Manuscript submitted by invitation to \emph{Annual Review of Biophysics})

\vspace{1cm}
\noindent
Daniel M. Zuckerman\\
Department of Computational and Systems Biology\\
University of Pittsburgh\\
ddmmzz@pitt.edu\\
(Corresponding author)

\vspace{1cm}
\noindent
Running title: Computer Sampling of Biomolecules



\section*{Abstract}
Equilibrium sampling of biomolecules remains an unmet challenge after more than 30 years of atomistic simulation.
Efforts to enhance sampling capability, which are reviewed here, range from the development of new algorithms to parallelization to novel uses of hardware.
Special focus is placed on classifying algorithms --- most of which are underpinned by  a few key ideas --- in order to understand their fundamental strengths and limitations.
Although algorithms have proliferated, progress resulting from novel hardware use appears to be more clear-cut than from algorithms alone, partly due to the lack of widely used sampling measures.

\section{Introduction}

\subsection{Why sample?}
Biomolecular behavior can be substantially characterized by the states of the system of interest --- that is, by the configurational energy basins reflecting the coordinates of all constituent molecules in a system. 
These states might be largely internal to a single macromolecule such as a protein, or more generally involve binding partners and their relative coordinates.
But regardless of the complexity of a system, the states represent key functional configurations along with potential intermediates for transitioning among the major states.
\underline{The primary goal of equilibrium computer simulation is to specify configura-} \underline{tional states and their populations.}
A more complete, mechanistic description would also include kinetic properties and dynamical pathways \cite{Zuckerman2010}.
Nevertheless, even the equilibrium description provides a basic view of the range of structural motions of biomolecules, and can also serve more immediately practical purposes such as for ensemble docking \cite{Oshiro-1997,Lin2002}.

The basic algorithm of biomolecular simulation, molecular dynamics (MD) simulation, has not changed substantially since the first MD study of a protein more than 30 years ago \cite{McCammon1977}. 
Although routine explicit-solvent MD simulations are now four or five orders of magnitude longer (i.e., $100$ - $10^3$ nsec currently), modern MD studies still appear to fall significantly short of what is needed for statistically valid equilibrium simulation (e.g., \cite{Grossfield2007a,Grossfield2009}).
Roughly speaking, one would like to run a simulation at least 10 times longer than the slowest important timescale in a system.
Unfortunately, many biomolecular timescales exceed 1 ms, and in some cases by orders of magnitude (e.g., \cite{Howard2001}).

Despite the outpouring of algorithmic ideas over the past decades, MD largely remains the tool of choice for biomolecular simulations.
While this staying power partly reflects the ready availability of software packages, and perhaps some psychological inertia, it also is indicative of a simple fact: no other method can routinely and reliably ``beat'' MD by a significant amount. 

This article will employ several points of view in considering efforts to improve sampling.
First, it will attempt to define ``the equilibrium sampling problem(s)'' as precisely as possible, which necessarily includes discussing the quantification of sampling.
Significant and substantiated progress is not possible without clear yardsticks.
Secondly, a brief discussion of how sampling happens --- i.e., how simulations generate ensembles and averages --- will provide a basis for understanding numerous methods.
Third, the article will attempt to review a fairly wide array of modern algorithms --- most of which are underpinned by a surprisingly small number of key ideas. 
The algorithms and studies are too numerous to be reviewed on a case-by-case basis, but a bird's eye perspective is very informative; more focused attention will be paid to apparently conflicting statements about the replica exchange approach, however. 
Fourth, special emphasis will be placed on novel uses of hardware --- GPUs, RAM, and special CPUs; this new ``front'' in sampling efforts has yielded rather clean results in some cases.
Using hardware in new ways necessarily involves substantial algorithm efforts.

This review is limited.
At least a full volume would be required to comprehensively describe sampling methods and results for biomolecular systems.
This review therefore aims primarily to catalog key ideas and principles of sampling, along with enough references for the reader to delve more deeply into areas of interest.
The author apologizes for the numerous studies which have not been mentioned for reasons of space or because he was not aware of them.
Other review articles (e.g., \cite{Straub-1997b,Christen2008}) and books on simulation methods \cite{Allen-Tildesley,Frenkel-book,Leach2001,Schlick2002} are available.

\section{What is the sampling problem?}
While scientists actively working in the field of biomolecular simulation may assume the essential meaning of ``the sampling problem'' is universally accepted, a survey of the literature indicates that several somewhat different interpretations are implicitly assumed.
What is meant by the sampling problem and, accordingly, success or failure in addressing the problem, surely will dictate the choice of methodology.

\subsection{A simple view of the equilibrium sampling problem}
Equilibrium sampling at constant temperature and fixed volume can be concisely be identified with the generation of full-system configurations $\xbf$ distributed according to the Boltzmann-factor distribution:
\begin{equation}
\label{boltz}
\rho(\xbf) \propto \exp \lb \left. -U \lp \xbf \rp \right/ k_B T \rb,
\end{equation}
where $\rho$ is the probability density function, $U$ is the potential energy function, $k_B$ is Botlzmann's constant, and $T$ is the absolute temperature in Kelvin units.
The coordinates $\xbf$ refer (in the classical picture assumed here) to the coordinates of every atom in the system, including solute and solvent.
Note that other thermodynamic conditions, such as constant pressure or constant chemical potential, require additional energy-like terms in the Boltzmann factor \cite{\dmzbook}, but here we shall consider only the distribution (\ref{boltz}) for clarity and simplicity. 

Performing equilibrium sampling requires access to all regions of configuration space, or at least to those regions with significant populations, and also requires that configurations have the correct relative probabilities.
This point will be elaborated on, below.

\subsection{Ideal sampling as a reference point}
\label{sec:ideal}
In practice, it is nearly impossible to generate completely independent and identically distributed (i.i.d.) configurations --- obeying \eq{boltz} --- for biomolecules.
Both dynamical and non-dynamical methods tend to produce correlated samples \cite{Mountain1989,Hess2002,Lyman2007a,Grossfield2009,Zhang2010}.
Nevertheless, such ideal sampling is a highly useful reference point for disentangling complications arising in practical simulation methods which generate correlated configurations.

\begin{figure}[h]
\begin{center}
\includegraphics[totalheight=2.5in]{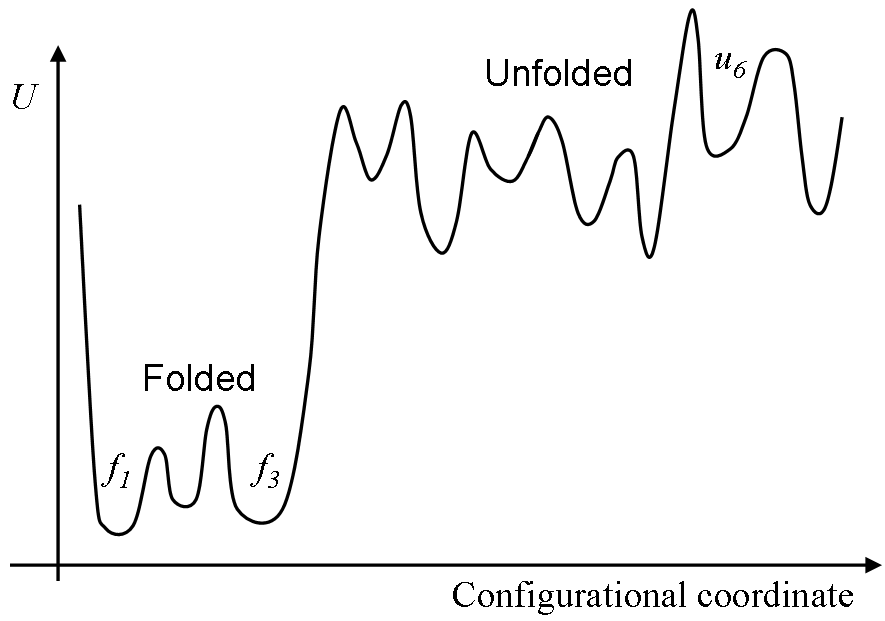}
\end{center}
\caption{
\label{fig:landscape}
A schematic energy landscape of a protein.
Both the folded states and unfolded states generally can be expected to consist of multiple sub-states, $f_i$ and $u_i$, respectively.
Transitions among sub-states themselves could be slow, compared to simulation timescales.
The ``sampling problem'' is sometimes conceptualized to involved generating configurations from both folded and unfolded states: the text discusses when this view is justified.
}
\end{figure}

Consider a hypothetical example where an equilibrium ensemble of $N$ i.i.d. configurations has been generated according to \eq{boltz} for a biomolecule with a complex landscape.
Regions of configuration space with probability $p \gtrsim 1/N$ are likely to be represented in the ensemble, and this is true regardless of kinetic barriers among states.
Conversely, regions with $p \lesssim 1/N$ are not likely to be represented, regardless of whether the region is ``interesting.''
For instance, the unfolded state might or might not be an appropriate part of a size $N$ ensemble, depending on the system (i.e., on the free energy difference between folded and unfolded states) and on $N$.
See Fig.\ \ref{fig:landscape} and Sec.\ \ref{sec:initial}, below.

\subsection{Goals and ambiguities in equilibrium sampling}
\label{sec:ambig}
The question, ``What should be achieved in equilibrium sampling?'' can be divided into at least three questions.
First, for a given problem (defined by the specific system and a specified initial condition), there is the question of defining success --- When is sampling sufficient?
Usually, the goal of equilibrium simulation is to calculate observables of interest to a level of precision sufficient to draw physical conclusions.
Yet the variables ``of interest'' will vary from study to study.
It therefore seems more reasonable to set a goal of generating an ensemble of configurations from which arbitrary observables can be calculated.
In rough terms, let us therefore assume that the goal is to calculate $\neff$ effectively independent configurations, where presumably it is desirable to achieve $\neff > 10$.
Larger values may often be necessary, however, because fractional uncertainty will vary as $1/\sqrt{\neff}$ for slower-to-converge properties such as state populations.
This issue will be addressed further below.

The other questions are probably the most pressing, starting with, ``What type of system is to be sampled?''
A basic distinction that seems to appear implicitly in the literature is whether the unfolded state is important in the system or not.
In this context, ``importance'' would usually be defined by whether or not a system exhibits a significant equilibrium unfolded fraction --- say, roughly equal probabilities $\pfold \sim \punf$.
Indeed, some evaluations of sampling methods have specifically targeted such a balance \cite{Huang2008,Rosta2009b}, e.g., by adjusting the temperature.

A third question about initial conditions is addressed separately, below, in Sec.\ \ref{sec:initial}.

These questions arise in the first place because all available practical sampling methods for biomolecules are ``imperfect'' in the sense of producing correlated configurations.
In ideal equilibrium sampling (Sec.\ \ref{sec:ideal}), by contrast, the sufficiency of sampling is fully quantified in terms of the number $N$ of independent configurations, and can be assessed objectively in terms of observables of interest.
The notion of an initial condition is irrelevant in ideal sampling since there are no correlations.
Also, the values of $\pfold$ and $\punf$ have no effect on sampling quality, which is fully embodied in $N$; instead, it is $N$ which determines what states are represented in the ensemble and the appropriate frequencies of occurrence.

\subsubsection{Should initial conditions matter? Issues surrounding the unfolded state}
\label{sec:initial}

A typical simulation of a protein can be described as one intended to sampled the folded state, which may consist of a number of substates as in Fig.\ \ref{fig:landscape}.
So long as the simulation is initiated in or near one of these substates, the resources required to sample it will not depend significantly on the initial conditions.
Rather, timescales for transitioning among substates will dominate the computing cost.
In a dynamical simulation, after all, it is the transitions among substates which yields estimates for substate populations and hence the equilibrium ensemble: see Sec.\ \ref{sec:transn}.

Sensitivity to initial conditions can become important when the unfolded state is involved.
Even when the goal is to sample the equilibrium ensemble of an overwhelmingly folded protein, the initial condition can come into play if it introduces an anomalously long timescale.
Consider beginning an equilibrium simulation of the landscape of Fig.\ \ref{fig:landscape} in the unfolded substate labeled $u_6$.
Depending on details of the system, the time required to find the folded state could easily exceed the time for sampling folded substates.
Thus, such a simulation, first has to solve a \underline{search} problem before sampling is begun.



We can set our discussion in the context the folding free energy, defined based on the ratio of probabilities for unfolded and folded states:
\begin{equation}
\label{gunf}
\pfold / \punf = \exp \lp \left. -\dgfold \right / k_B T \rp \; .
\end{equation}
Consider a value $\dgfold \sim -3 \mbox{kcal/mol} \sim -5 k_B T$, which implies the unfolded state is occupied 1\% of the time or less.
In ensembles with fewer than 100 \underline{independent} configurations ($\neff < 100$), such an unfolded state typically should not be represented.
In practical terms, because key \underline{folded-state} timescales often exceed $\mu$sec or even msec, it is unlikely that a simulation of a protein in folding conditions in the current era will contain a sufficient number of independent configurations for the unfolded state to be represented.

Note that we have assumed it is straightforward to distinguish folded and unfolded configurations.
In reality, there may be a spectrum of disordered states in many systems, and hence ambiguity in defining folded and unfolded states.

\section{Sampling basics: Mechanism, timescales, and cost}

\subsection{Transitions provide the key information in sampling}
\label{sec:transn}

The problem of equilibrium sampling can be summarized as determining the metastable states and their relative populations, but how are populations actually determined by the simulation?
In almost every practical algorithm, the information comes from \underline{transitions among} \underline{the states.}
By switching back and forth between pairs of states, with dwell times between transitions acting as proxies for rate constants $k$, a dynamical simulation gathers information about relative populations based on the equilibrium balance relation $p_i k_{ij} = p_j k_{ji}$ \cite{\dmzbook}.
With more transitions, the error in the population estimates $p_i$ decreases \cite{Rosta2009b}. 
Without transitions, most algorithms have no information about relative populations.
Consider the case of two independent simulations started from different states which exhibit no transitions: determining the populations of the states then requires more advanced analysis \cite{Ytreberg2008} which often may not be practical.

It is critical realize that most ``advanced algorithms'' --- such as the varieties of exchange simulation (Sec.\ \ref{sec:algo}) --- also require ``ordinary'' transitions to gather population information.
That is, transitions among states within individual continuous trajectories are required to obtain sampling, as has been discussed before \cite{Zheng2007a,Grossfield2009,Rosta2009b}. 
Any algorithm that uses dynamical trajectories as a component can be expected to require transitions among metastable states in those trajectories.

It must also be pointed out that \underline{transitions themselves do not necessarily equate to} \underline{good sampling}.
An example is when a high temperature is used to aid sampling at a lower temperature.
Although transitions are necessary, they are not sufficient because of the overlap issue: the sampled configurations ultimately must be important in the targeted (e.g., lower temperature) ensemble after they are reweighted \cite{Lyman2007,Shen2008}.

\subsection{Timescales of sampling}
\label{sec:times}
Much of the foregoing discussion can be summed up based on two key timescales.
(Even if a sampling method is not fully dynamical, analogous quantities can be defined in terms of computing times.)
The timescale that typically is limiting for simulations can be denoted as $\tcstar$ which denotes the longest ``important'' equilibrium correlation time.
Roughly, this is the time required to explore all the important parts of configuration space \underline{once} \cite{Lyman2007a} --- starting from any well populated (sub)state --- and is elaborated on below.
For sampling the landscape of Fig.\ \ref{fig:landscape} under folding conditions, this would be the time to visit all folded substates starting from any of the folded basins. 
However, as also discussed above, initial conditions may play an important role, and thus we can define $\tinit$ as the ``equilibration'' time (or ``burn-in'' time in Monte Carlo lingo).
This is \underline{not} an equilibrium timescale, but the time necessary to relax from a particular non-equilibrium initial condition to equilibrium.
Thus, $\tinit$ is specific to the initial configuration of each simulation.
The landscape of Fig.\ \ref{fig:landscape} under folding conditions suggests $\tinit \ll \tcstar$ for a simulation started from a folded substate, but it is possible to have significant values --- even $\tinit > \tcstar$ --- when starting from an unfolded state.

\subsection{Factors contributing to sampling cost} 
\label{sec:think}
A little thought about the ingredients of sampling can help one understand successes and failure of various efforts --- and more importantly, aid in planning future research.
The most important ingredient to understand is single-trajectory sampling --- i.e., the class of dynamical methods discussed in Sec.\ \ref{sec:dynam} including MD, single-Markov-chain Monte Carlo (MC), or any other method that generates an ensemble as a sequentially correlated list of configurations. 
Single-trajectory methods tend to be the engine underlying traditional and advanced methods. 

The sampling cost of a trajectory method can be divided into two intuitive factors:
\begin{eqnarray}
\label{trajcost}
\mbox{Trajectory sampling cost} & =  &
  \mbox{(Cost per trajectory step)}  \times \mbox{(Number of steps needed for sampling)}
  \nonumber \\
  & \sim & \mbox{(Cost of energy call)} \times \mbox{(Roughness of energy landscape)} \; .
\end{eqnarray} 
In words, a given algorithm will require a certain number of steps to achieve good sampling (e.g., $\neff \gg 1$) and the total cost of generating such a trajectory is simply the total cost per step multiplied by the required number of steps.
Although this decomposition is trivial, it is very informative.
For example, if one wants to use a single-trajectory algorithm (e.g., MD or similar) where a step corresponds roughly to $10^{-15}$ sec, then $10^{9}$ - $10^{12}$ steps are required to reach the $\mu$sec - msec range.
Therefore, unless the cost of a step can be reduced by several orders of magnitude, MD and similar methods likely cannot achieve sampling with typical current resources.
In a modified landscape (e.g., different temperature or model) the roughness may be reduced --- but, again, that reduction should be several orders of magnitude, or else be accompanied by a compensating reduction in the cost per step.
Among noteworthy examples, increasing temperature decreases roughness but not the cost per step, whereas changing a model (i.e., modifying the potential energy function $U$) can affect both roughness and the step cost.
Both strategies sample a different distribution, which may be nontrivial to convert to the targeted ensemble.

The importance of understanding dynamical trajectory methods via \eq{trajcost} is underscored by the fact that most multi-level (e.g., exchange) simulations require good sampling at some level \cite{Zuckerman2006,Zheng2007a}, by single continuous trajectories.

\section{Quantitative assessment of sampling}
\label{sec:ess}
Although the challenge of sampling biomolecular systems has been widely recognized for years, and although new algorithms are regularly proposed, systematic quantification of sampling has often not received sufficient attention.
Most importantly, there does not seem to be a widely accepted (and widely used) ``yardstick'' for quantifying the effectiveness of sampling procedures.
Given the lack of a broadly accepted measure, it is tempting for individual investigators to report measures which cast the most favorable light on their results.

Two key benefits would flow from a universal measure of sampling quality.
As an example, assume we are able to assign an effective sample size, $\neff$, to any ensemble generated in a simulation; $\neff$ would characterize the number of statistically independent configurations generated.
Sampling efficiency could then be quantified by the CPU time (or number of cycles) required per independent sample.
The first benefit is that we would no longer be able to fool ourselves or others as to the efficiency of a method based on qualitative evidence or perhaps the elegance of a method.
Thus, the field would be pushed harder to focus on a bottom-line measure.
Secondly, describing the importance of a new algorithm would be more straightforward.
It would no longer be necessary to directly compare a new method to, perhaps, a competitor's approach requiring subtle optimization.
Instead, each method could be compared to a standard (e.g., molecular dynamics), removing ambiguities in the outcome.
It would still be important to test a method on a number of systems (e.g., small and large, stiff and flexible, implicitly and explicitly solvated), but the results would be quantitative and readily verified by other groups.
On a related note, there has been a proposal to organize a sampling contest or challenge event to allow head-to-head comparison of different methodologies (B.R.\ Brooks, unpublished).


Numerous ideas for assessing sampling have been proposed over the years (e.g., \cite{\assessbrief}), but it is important to divide such proposals into ``absolute'' and ``relative'' measures.
Absolute measures attempt to give a binary indication of whether or not ``convergence'' has been achieved, whereas relative measures estimate how much sampling has been achieved --- e.g., by an effective sample size $\neff$.
The perspective of the present article is that absolute measures fail to account for the fundamental statistical picture underlying the sampling problem.
To see why, note that any measured observable will have an uncertainty associated with it; roughly speaking, sampling quality is reflected in typical sizes of error bars which decrease with better sampling.
There does not seem to an unambiguous point where sampling can be considered absolutely converged (although a simulation effectively does not begin to sample equilibrium averages until $\tinit$ has been surpassed). 

It is certainly valuable, nevertheless, to be able to gauge when sampling is wholly inadequate, in an absolute sense.
Absolute methods which may be useful at detecting extremely poor sampling include casual use of the ``ergodic measure'' \cite{Mountain1989} (i.e., whether or not it approaches zero) and cluster counting \cite{vanGunsteren-2002}.

\subsection{Assessing dynamical sampling}
\label{sec:dynam}

The assessment of dynamical methods illustrates the key ideas behind the \underline{relative} measures of sampling.
For this purpose, ``dynamical'' will be defined to mean when trajectories consist of configurations with purely sequential correlations --- i.e., where a given configuration was produced based solely on the immediately preceding configuration(s).
Thus, MD, Langevin dynamics, and simple Markov-chain MC are dynamical methods, but exchange algorithms are not \cite{\dmzbook}.
Correlation times or their analogs for MC can then readily be associated with any trajectory which was dynamically generated.
Assuming, for the moment, that there is a fundamental correlation time for overall sampling of the system, $\tcstar$, then an effective sample size can be calculated from the simple relation
\begin{equation}
\label{neffdyn}
\neff \simeq \tsim \left/ \tcstar \right. \; ,
\end{equation}
where $\tsim$ is the total CPU time (or number of cycles) used to generate the trajectory.
Thus, sampling quality is \underline{relative} in that it should increase linearly with $\tsim$ \cite{Mountain1989}.
In rough terms, the \underline{absolute} lack of sampling would correspond to $\neff \lesssim 1$; physically, this would indicate that important parts of configuration have been visited only once --- e.g., a simulation shorter than $\tinit + \tcstar$.
(It would seem impossible to detect parts of the space \underline{never} visited.)

Recent work by Lyman and Zuckerman has suggested that a reasonable overall correlation time $\tcstar$ can be calculated from a dynamical trajectory \cite{Lyman2007a}.
Those authors derived their correlation time from the overall distribution in configuration space, estimating the time that must elapse between trajectory frames so that they behave as if statistically independent.
The approach has the twin advantages of being based on the full configuration-space distribution (as opposed to isolated observables) and of being blindly and objectively applicable to any dynamical trajectory.
Other measures meeting these criteria would also be valuable; note, for instance, the work by Hess using principal components \cite{Hess2002}.

\subsection{Assessing non-dynamical sampling}

If sampling is performed in a non-dynamical way, one cannot rely on sequential correlations to assess sampling as in \eq{neffdyn}.
Many of the modern algorithms which attempt to enhance sampling, such as those reviewed below, are not dynamical.
Replica exchange (RE) \cite{\repexalgo} is a typical example: if one is solely interested in the ensemble at a single temperature, a given configuration may be strongly correlated with other configurations distant in the sequence at that $T$ and/or be uncorrelated with sequential neighbors.
At the same time, RE does generate trajectories which are continuous in configuration space, if not temperature, and it may be possible to analyze these in a dynamical sense \cite{Lyman2007a} but care will be required \cite{Chodera2007a}.  
Other algorithms, for instance based on polymer growth procedures \cite{Velikson1992,Grassberger1997,Frauenkron1998,Zhang2002,Zhang2007b,Zhang2008a}, are explicitly non-dynamical.

Two recent papers \cite{\GrossfieldZuckerman,Zhang2010} have argued that non-dynamical simulations are best assessed by multiple independent runs.
The lack of sequential correlations --- but the presence of more complex correlations --- in non-dynamical ensembles means that the list of configurations cannot be divided into nearly independent segments for blocking-based analysis. 
Zhang et al.\ suggest that multiple runs be used to assess variance in the populations of physical states \cite{Zhang2010} for two related reasons: (i) such states are defined to be separated from one another by the slowest timescales in a system;
(ii) relative populations of states cannot be properly estimated without good sampling within each state.
State-population variances, in turn, can be used to estimate $\neff$ based on simple statistical arguments \cite{Zhang2010};
nevertheless, it is important that states be approximated in an automated fashion --- see \cite{Chodera2007,Noe2007,Zhang2010} --- to eliminate the possibility for bias.

\section{Purely algorithmic efforts to improve sampling}
\label{sec:algo}

How can we beat MD?
Despite 30 years of effort, there is no algorithm that is significantly more efficient than molecular dynamics for the full range of systems of interest.
Further, although dozens of different detailed procedures have been suggested, there are a limited number of qualitatively distinct ideas.
At attempt will now be made to describe some of the strategies which have been proposed.

\subsection{Replica exchange and multiple temperatures}
\label{sec:repex}
\begin{figure}[h]
\begin{center}
\includegraphics[totalheight=1.75in]{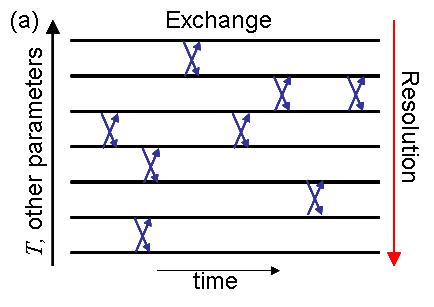}
\includegraphics[totalheight=1.75in]{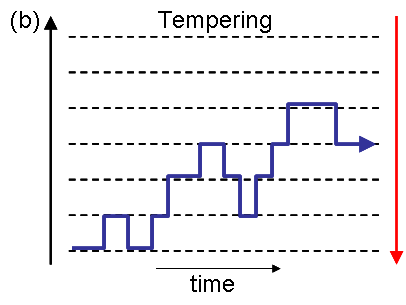}
\includegraphics[totalheight=1.75in]{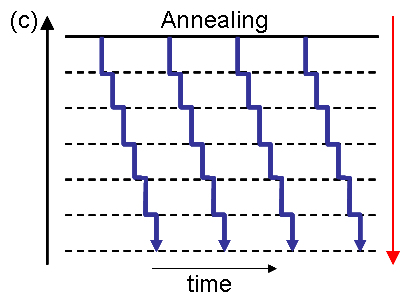}
\end{center}
\caption{
\label{fig:ladders}
Many ways to use many levels.
Different horizontal levels represent different simulation conditions, such as varying temperature, model parameters, and/or resolution.
A given ladder or set of conditions can be used in different formulations: (a) exchange, (b) tempering, or (c) annealing.
}
\end{figure}

The most common strategy is to employ elevated temperature.
Many variations on this strategy have been proposed, with one of the earliest suggestions being in the context of spin systems \cite{Swendsen-1988}.
The approach that has been applied to biomolecules most often is replica exchange (RE, also called parallel tempering) in which parallel simulations wander among a set of fixed temperature values with swaps governed by a Metropolis criterion \cite{\repexalgo}.
Many variations and optimizations have been proposed for RE (e.g., \cite{\repexvar,\repexopt}).
Closely related to RE is simulated tempering, in which a single trajectory wanders among a set of temperatures \cite{Lyubartsev1992}; once again, optimizations have been proposed (e.g., \cite{Zhang2010c}).
See Fig.\ \ref{fig:ladders}.

Less well known, but formally closely related to exchange and tempering, is ``annealed importance sampling'' (AIS), in which a high-temperature ensemble is annealed to lower temperatures in a weighted way that preserves canonical sampling \cite{\aisgen}.
See the schematic representation in Fig.\ \ref{fig:ladders}.
AIS has been applied to biomolecules and optimizations have been suggested \cite{\aisdmz}.
AIS is nominally a non-equilibrium approach, but in precise analogy to the Jarzynski equality \cite{Jarzynski-1997a}, it yields equilibrium ensembles.
The ``J-walking'' approach also starts from high temperature \cite{Doll-1990}; a good discussion of several related methods is given in Ref.\ \cite{Head-Gordon-2002}.

\subsubsection{How effective is replica exchange?}
Replica exchange (RE) simulation is both popular and controversial.
Some authors have noted weaknesses \cite{\repexcrit}, others have described successes (e.g., \cite{Huang2008,Sanbonmatsu2002,Rosta2009b}), and some conclusions have been more ambiguous \cite{Periole2007a,Beck2007,Zheng2007a,Machta2009}.
What is the real story --- how much better is RE than ordinary MD?

Examination of the various claims and studies reveals that, in fact, there is little disagreement \underline{so long as the particular sampling problem is explicitly accounted for (Sec.\ \ref{sec:ambig}).}
In brief, when observables of interest depend significantly on states which are very rapidly accessed at high temperature, then replica exchange and related methods can be very efficient \cite{Huang2008,Rosta2009b}.
A prime example would be estimating the folded fraction near the folding transition.
On the other hand, when the goal is to sample an overwhelmingly folded ensemble, there does not appear to be significant evidence of replica exchange's efficiency --- so long as the initial condition is a folded configuration (i.e., in terms of the timescales of Sec.\ \ref{sec:times}, if $\tinit < \tcstar$).
This last qualification is important in understanding the data presented in Ref.\ \cite{Huang2008}.

An intuitive picture underpinning the preceding conclusions is readily gained by re-visiting the idea that state populations are estimated by transitions between states in continuous trajectories (Sec.\ \ref{sec:transn}).
Indeed, this same picture is the basis for discrete-state analyses \cite{Zheng2007a,Rosta2009b}, which we can consider along with the schematic landscape of Fig.\ \ref{fig:landscape}.
Specifically, the average of any observable $A$ can be written as $\lav A \rav = \sum_i p_i \bar{A}_i$, where $i$ indexes all folded and unfolded substates and $\bar{A}_i$ is the average of $A$ within substate $i$.
If $p_i$ values are significant in both folded and unfolded substates, then the elevated temperatures employed in RE simulation can usefully promote transitions to sample unfolded states \cite{Rosta2009b}.
On the other hand, if $\punf \ll \pfold$, where $\pfold = \sum^{\mathrm{folded}}_i p_i$ is the sum of folded probabilities and $\punf = 1 - \pfold$, then computer resources devoted to unfolded substates can detract from estimating folded-state observables; this line of analysis echoes earlier studies \cite{Zheng2007a,Nymeyer2008}.

Regardless of whether RE is superior to MD, another key issue is whether RE can provide \underline{sufficient} \underline{sampling} ($\neff > 10$) for protein-sized systems.
It is far from clear that this is the case.
RE cannot provide full sampling unless some level of the ladder can be fully sampled \cite{Zuckerman2006} (see Sec.\ \ref{sec:transn}), and the ability for a simple dynamical trajectory to sample the large configuration space of a fully or partly unfolded state of a protein has not been demonstrated to date.

\subsubsection{Energy-based sampling methods}
Energy-based schemes must be seen as closely related to temperature-based strategies since the two variables are conjugate to each other in the Boltzmann factor, \eq{boltz}.
Yet because any fixed-temperature ensemble can include arbitrary energy values, energy-based schemes must account for this.  
``Multi-canonical'' schemes based on sampling energy values (typically uniformly) have been implemented for biomolecules \cite{\multicanon}, including with the Wang-Landau device \cite{Wang2001}; see also \cite{Kim2006}.
Nevertheless, it should be remembered that energy is not expected to be a good proxy for the configurational coordinates of primary interest --- see Fig.\ \ref{fig:landscape}. 

\subsection{Hamiltonian exchange and multiple models}
\label{sec:hamex}

As schematized in Fig.\ \ref{fig:ladders}, the different temperature-based methods sketched above can readily be generalized to variations in forcefield parameters \cite{Okamoto-2000}, and even resolution as discussed below.
This is because the methods are based on the Boltzmann factor, which is a general form that can apply to different forcefields $\ulam$, each characterized by the set of parameters $\lamvec = \{\lambda_1, \lambda_2, \ldots \}$.
Thus, the fundamental distribution \eq{boltz} can be written more explicitly as:
\begin{equation}
\label{boltzgen}
\rho(\xbf; T, \lamvec) \propto \exp \lb \left. -\ulam \lp \xbf \rp \right/ k_B T \rb,
\end{equation}
The different schemes for using ladders of different types --- based on temperature, forcefield parameters, or resolution --- are depicted in Fig.\ \ref{fig:ladders}.
One of the first proposals for using multiple models was given in Ref.\ \cite{Liu1993}.

The simplest way to reduce the roughness of a landscape (cf.\ \eq{trajcost}) is by maintaining the functional form of $U$ but changing parameters --- e.g., coefficients of dihedral terms.
Changing parameters in parameter files of common software packages makes this route fairly straightforward. 
Note that unless forcefield terms are explicitly removed, the step cost in \eq{trajcost} remains unchanged. 

Hamiltonian exchange has been applied in a number of cases, for instance using models with softened van der Waals' interactions and hence decreased roughness \cite{Hritz2008a}.
The approach called ``accelerated MD'' employs ordinary MD on a smoothed potential energy landscape and requires reweighting to obtain canonical averages \cite{Hamelberg2004a,Shen2008}; it can therefore be considered a two-level Hamiltonian-changing algorithm.

\subsection{Resolution exchange and multi-resolution approaches}
\label{sec:resex}
Extending the multiple-model ideas a step further leads to the consideration of multiple resolutions. 
To account for varying resolution in the formulation of \eq{boltzgen}, some of the $\lambda_i$ parameters may be considered prefactors which can eliminate detailed interactions when set to zero.
Different, formally exact approaches to using multiple resolutions have been proposed \cite{\resexdmz,\resexvar}, primarily echoing ideas in replica exchange and annealing.

Although multiple-resolution methods have not produced dramatic results for canonical sampling of large atomistic systems, they appear to hold unique potential in the context of the sampling costs embodied in \eq{trajcost}.
In particular, low-resolutions models reduce cost and roughness to such an extent that some coarse models have been shown to permit good sampling for proteins ($\neff > 10$) with typical resources \cite{Ytreberg2007,Mamonov2009b}.
It is not clear that a similar claim can be made for high-temperature or modified-potential simulations of stably folded proteins.

\subsection{Multiple trajectories \underline{without} multiple levels}
There is a qualitatively different group of algorithms that also uses multiple trajectories, but \underline{all} under the same conditions --- i.e., ``uni-canonically'' \cite{Huang2009,Noe2009,Bhatt2010a} --- in contrast to exchange and related methods.
The basic idea of these methods is that when a trajectory reaches a new region of configuration space, it can spawn multiple daughter trajectories, all on the unaltered landscape.
The daughter trajectories enable better sampling of regions that otherwise would receive less computer resources --- and also, resources can be directed away from regions which already are well sampled.
In the context of Markov-state models, this strategy can be optimized to increase precision in the estimate of any observable \cite{Hinrichs2007}.
Exact canonical sampling can also be using a steady-state formulation of the ``weighted ensemble'' path sampling method \cite{Bhatt2010a}; see also related approaches \cite{Raiteri2006,Warmflash2007,Valeriani2007}.

\subsection{Single-trajectory approaches: Dynamics, Monte Carlo, and variants}
It is useful to group together a large group of approaches --- including MD, simple Monte Carlo (MC) and Langevin dynamics \cite{Allen-Tildesley,Frenkel-book,Schlick2002} --- and consider them as basic dynamics \cite{\dmzbook}.
These methods generate a single trajectory (or Markov chain) in which configurations are strongly sequentially correlated --- and have no non-sequential correlations.
In other words, these methods should all be expected to behave roughly like MD simulation because the motion is intrinsically constrained by the high density and landscape complexity of biomolecular systems.
On the one hand, basic dynamical methods will generate physically realizable trajectories or nearly so, but on the other, they will be severely limited by physical timescales as described above.

In this section, we will explore some basic ideas of dynamical approaches and also some interesting variants. 

\subsubsection{Are there ``real'' dynamics?}
Unlike many other sampling algorithms, MD also simulates the fundamental classical dynamics of a system.
That is, the trajectories produced by MD are also intended to model the time-dependence of physical processes.

Should MD trajectories provide much better depictions of the underlying processes than other ``basic'' simulation methods --- e.g., Langevin or MC simulation?
The answer is not so clear.
First, like other methods, MD necessarily uses an approximate forcefield (see, e.g., \cite{Freddolino2009}).
Second, on the long timescales ultimately of interest, roundoff errors can be expected to lead to significant deviations from the exact trajectories for the given forcefield.

Somewhat more fundamentally, there appears to be a physical inconsistency in running finite-temperature MD simulations for finite-sized systems.
Thermostats of various types are used \cite{Frenkel-book}, some stochastic and some deterministic: is there a correct method?
By definition, a finite-temperature system is not isolated but is coupled to a ``bath'' by some physical process, presumably molecular collisions.
Because the internal degrees of freedom of a bath are not explicitly simulated, again by definition, a bath should be intrinsically stochastic.
The author is unaware of a first-principles prescription for modeling the coupling to a thermal bath, although sophisticated thermostats have been proposed \cite{Leimkuhler2004,Frenkel-book}.
For MD simulations with periodic boundary conditions, it is particularly difficulty to imagine truly physical coupling to a bath.

The bottom line is that a finite-temperature process in a finite system is intrinsically stochastic, rendering questionable the notion that standard MD protocols produce ``correct'' trajectories. 
To consider this point another way, should we really expect that inertial effects are very important in dense aqueous and macromolecular media?

The foregoing discussion suggests, certainly, that Langevin simulations should also be considered physical, but what about Monte Carlo?
Certainly, one can readily imagine unphysical MC protocols (e.g., attempting trial moves for atoms in the C-terminal domain of a protein ten times as often as for those in the N terminus). 
Nevertheless, so long as trial moves are small and performed in a spatially uniform way, one can expect MC ``dynamics'' to be an approximation to over-damped (non-inertial) Langevin simulation; see \cite{\mcdyn}. 

\subsubsection{Modifying dynamics to improve sampling}
\label{sec:moddyn}

Although changing temperature or model parameters automatically changes the dynamics observed in MD or LD simulation, it has been shown that the dynamics can be modified and maintain canonical sampling on the original landscape, $U(\xbf)$ \cite{Zhu2002,Minary2004a}.
This possibility defies the simple formulation of the second line of \eq{trajcost}, because the landscape roughness remains unchanged while the number of required steps is reduced.
The method has been applied to alkane and small-peptide systems, but its generalization to arbitrary systems has not been presented to the author's knowledge.
A better known and related approach is the use of multiple time steps, with shorter steps for faster degrees of freedom \cite{Teleman1986,Tuckerman1992}, which can save computer time up to a limit set by resonance phenomena \cite{Schlick1998}. 

Another approach to modifying dynamics is to avoid revisiting regions already sampled by means of a ``memory'' potential which repels the trajectory from prior configurations \cite{Huber1994,Laio2002}.
This strategy is called ``metadynamics'' or the local-elevation method, and typically employs pre-selection of important coordinates to be assigned repulsive terms.
Although trajectories generated in a history-dependent way do not satisfy detailed balance, it is possible to correct for the bias introduced and recover canonical sampling \cite{Bussi2006a}.
This approach could also be considered a potential-of-mean-force method (Sec.\ \ref{sec:pmf}).

The potential surface and dynamics can also be modified using a strategy of raising basins while keeping barriers intact \cite{Voter1997,Rahman2002,Hamelberg2004a}; see also \cite{Zuckerman-1999}.
The ``accelerated MD'' approach is a well-known example, but achieving canonical sampling requires a reweighting procedure \cite{Shen2008}.
Such reweighting is limited by the overlap issues confronting many methods: the sampled distribution must be sufficiently similar to the targeted distribution \eq{boltz} so that the data is not overwhelmed by statistical noise \cite{Lyman2007,Shen2008}.

Qualitatively, how can we think about the methods just described?
For one thing, as dynamical methods, they all gather information by transitions among states (Sec.\ \ref{sec:transn}): more transitions suggests greater statistical quality of the data.
However, methods which modify the potential surface may increase the number of transitions, but the overlap between the sampled and targeted distributions generally is expected decline with more substantial changes to the potential.
In this context, the modification of dynamics without perturbing the potential \cite{Zhu2002,Minary2004a} seems particularly intriguing, even though technical challenges in implementation may remain.

\subsubsection{Monte Carlo approaches}
The term ``Monte Carlo'' can be used in many ways, so it useful to first delimit our discussion.
Here, we want to focus on single-Markov-chain Monte Carlo, in which a sequence of configurations is generated, each one based on a trial move away from the previous configuration.
Typically, such MC simulations of biomolecules have a strong dynamical character because trial moves tend to be small and physically realizable in a small time increment \cite{\mcdyn}, whereas large trial moves would tend to be rejected in any reasonably detailed model.
As we shall see, however, less physical moves sometimes can be used --- including move sets which do not strictly obey detailed balance but instead conform to the weaker balance condition \cite{Manousiouthakis1999}.
It is also worth noting that the various exchange simulations discussed above (Secs.\ \ref{sec:repex}, \ref{sec:hamex}, \ref{sec:resex}) can be considered MC simulations because ``exchange'' is a special kind of trial move and indeed governed by a Metropolis acceptance criterion \cite{\dmzbook}.

A key advantage of single-chain MC simulation is that it can be used with any energy function, whether continuous or not, because forces do not need to be calculated \cite{Frenkel-book}.
MC simulation is available for use with standard forcefields in some simulation packages \cite{Jorgensen-1996b,Hu2006} but it seems to be most commonly used in connection with simplified models (e.g., \cite{Shimada2001,Zuckerman-2004d}).
The choice to use MC is usually for convenience: it readily enables the use of simple or discontinuous energy functions; furthermore, MC naturally handles constraints like fixed bond lengths and/or angles.
Substantial effort has gone into developing trial moves useful for biomolecules, with a focus on moves that act fully locally (e.g., \cite{Vendruscolo1997,Wu1999,Ulmschneider2003,Betancourt2005,Smith2008}).
Two studies of MC employing quasi-physical trial moves have reported efficiency gains compared to MD \cite{Jorgensen-1996b,Ulmschneider2006}.
MC has also been applied to large implicitly solvated peptides \cite{Mao2010}.
Further, it is widely used in non-statistical approaches such as docking \cite{Leach2001}, and it also is used in the Rosetta folding software \cite{Das2008}.

For implicitly solvated all-atom peptides, a novel class of trial moves has proven extremely efficient.
Ding et al.\ showed that when libraries of configurations for each amino acid were computed in advance, swap moves with library configurations provided extremely rapid motion in configuration space \cite{Ding2010}.
Efficiency gains based on measuring $\neff$ suggested the simulations were 100 - 1,000 times faster than MD.
Trial moves involving entire residues apparently were successful because subtle correlations among the atoms were accounted for in the pre-computed libraries; by comparison, internal-coordinate trial moves of a similar magnitude (e.g., twisting $\phi$ and $\psi$ dihedrals) were rarely accepted \cite{Ding2010}.

More detailed discussion of Monte Carlo methods can be found in recent reviews \cite{Earl2008,Vitalis2009} and textbooks \cite{Allen-Tildesley,Frenkel-book}. 

\subsection{Potential-of-mean-force methods}
\label{sec:pmf}
Many methods are designed to calculate a potential of mean force (PMF) for a specified set of coordinates $\chivec$, and such approaches implicitly are sampling methods.
After all, the Boltzmann factor of the PMF is defined to be proportional to the probability distribution for the specified coordinates: $\rho(\chivec) \propto \exp[-\pmf(\chivec) / k_B T ]$ \cite{\dmzbook}.
If the PMF has been calculated well (with sufficient sampling along the $\chivec$ coordinates \underline{and} orthogonal to them) then the configurations can be suitably reweighted into a canonical ensemble.

There are numerous approaches geared toward calculating a PMF. 
Some explore the full configuration space in a single trajectory, such as lambda dynamics \cite{\lambdadyn},  and could be categorized with modified dynamics methods, although this is just a semantic issue.
A large fraction of PMF calculations employ the weighted histogram analysis method (WHAM) \cite{Swendsen-1992}, but many alternatives have been developed (e.g., \cite{\pmfrefs}).

The main advantage which PMF methods hope to attain is faster sampling along the $\chivec$ coordinates than would be possible using brute-force sampling.
This perspective allows several observations.
First, to aid sampling, the investigator must be able to select \underline{all} important slow coordinates in advance --- otherwise, sampling in directions ``orthogonal'' to $\chivec$ will be impractical.
Second, the maximum advantage which can be gained over a brute-force simulation will depend on how substantial the barriers are along the $\chivec$ coordinates; if sampling along a coordinate is slow because many states are separated by only small barriers, then the advantage may be modest.
Last, for a PMF calculation to be successful, different values of the $\chivec$ coordinates must be quantitatively related to each other: 
depending on the details of the method, this can be directly via numerous transitions (e.g., \cite{\lambdadyn,Swendsen-2004,Zheng2008a}) echoing Sec.\ \ref{sec:transn}, or by requiring well-sampled sub-regions of overlap \cite{Swendsen-1992}.

\subsection{Non-dynamical methods}
Methods which do not rely on dynamics for sampling use a variety of distinct strategies, but as there are relatively few such efforts, they are grouped together for convenience.

Recently, there have been a number of applications to biomolecules of old polymer growth ideas (e.g., \cite{Wall1959,Meirovitch1982}), sometimes termed ``sequential importance sampling'' \cite{Liu2002}.
The basic idea is to add one monomer (e.g., amino acid) at a time to an ensemble of partially grown configurations while (i) keeping track of appropriate statistical weights and/or (ii) by ``resampling'' \cite{Liu2002}.
These approaches have been applied to simplified models of proteins and nucleic acids \cite{Frauenkron1998,Zhang2002,Zhang2007b,Zhang2008a}, to all-atom peptides at high temperature \cite{Velikson1992}, and more recently to all-atom peptides using pre-sampled libraries of amino-acid configurations \cite{Mamonov2009a}.
The intrinsic challenge in polymer growth methods is that configurations important in the full molecule may have low statistical probability in early stages of growth;
thus application to large, detailed systems likely will require biasing toward structural information known for the full molecule \cite{Zhang2007b,Mamonov2009a}.

Some methods have been developed which attempt a semi-enumerative description of energy basins.
For instance, the ``mining minima'' method uses a quasi-harmonic procedure to estimate the partition function of each basin, which in turn determines the relative probability of the basin \cite{AChang2007}; see also \cite{Bogdan2006}.
While such approaches have the disadvantage of requiring an exponentially large number of basins \cite{Wales2004}, they largely avoid timescale issues associated with dynamical methods because the basins are located with faster search methods.

Another class of approaches has attempted to treat the problem of estimating the free energy of a previously generated canonical ensemble \cite{Killian2007,Hnizdo2008,Ytreberg2008}; see also \cite{Meirovitch-2004b}.
In analogy to PMF methods (Sec.\ \ref{sec:pmf}), knowledge of free energies for independent simulations in separate states (for which no transitions have been observed) enables the states to be combined in a full ensemble: the free energies directly imply the relative probabilities of the states.  
The ideas behind such methods are fairly technical and beyond the scope of this review.

\section{Special hardware use for sampling}
\label{sec:hardware}
In several cases, rather impressive improvements in sampling speed have been reported based on using novel hardware --- and based on novel uses of ``ordinary'' hardware.
It is almost always the case that using new hardware also requires algorithmic development, and that component should not be minimized.
Little is ``plug and play'' in this business.

\subsubsection{Parallelization, special-purpose CPUs, and distributed computing} 

Perhaps the best known way to employ hardware is by parallelization.
On the one hand, parallelization typically makes sampling \underline{less} efficient than single-core simulation when measured by sampling per core (or per dollar) due to overhead costs.  
On the other, parallelization allows by far the fastest sampling for a given amount of wall-clock time (for a single system) including record-setting runs \cite{Duan1998,Freddolino2008,Shaw2009}.
Recent examples of parallelization of single-trajectory MD include: a 10 $\mu$sec simulation of a small domain \cite{Freddolino2008}; $\mu$sec+ simulations of membrane proteins on the BlueGene \cite{Grossfield2008} and Desmond \cite{Dror2009a} platforms; and the longest reported to date, a msec simulation of a small globular protein on the Anton machine \cite{Shaw2009}; see also \cite{Ensign2007}. 
The Anton simulation reflects both parallelization and the use of special-purpose chips, which can be inferred to each provide $\sim$10 or more times speedup compared to standard chips \cite{Shaw2007,Shaw2009}.

Long simulations have significant value.
They allow the community to study selected systems in great detail and to appreciate phenomena which could not otherwise be observed.
Of equal importance, long simulations can alert the community to limitations of MD and forcefields \cite{Grossfield2007a,Freddolino2009}.   

There are other parallel strategies, such as exchange simulation (see above) and distributed computing.
Distributed computing employs many simultaneous independent simulations or --- via repeated rounds of simulation --- quasi-parallel computing with minimal communication among simulations.
While distributed computing has primarily been applied to the folding problem \cite{Snow2002a}, recent work has shown the value of multiple short simulations for producing Markov state models \cite{Huang2009,Noe2009}.
Such models can be used to deduce both non-equilibrium and equilibrium information --- thus canonical ensembles if desired.
A distributed computing framework can also be used for multi-level simulations such as replica exchange \cite{\repexdistrib} and in principle for other quasi-parallel methods \cite{Bhatt2010a}.


\subsection{Use of GPUs and RAM for sampling}
There are other means to exploit hardware besides parallelization.
Combining a GPU (graphics processing unit) and CPU, implicit-solvent simulations of all-atom proteins can be performed hundreds of times faster than by a CPU alone \cite{Friedrichs2009}. 
Other work has exploited standard memory capacity (RAM) available on modern computers:
by pre-calculating ``libraries'' of amino acid configurations, ``library-based Monte Carlo'' was shown to sample implicitly solvated peptides hundreds and sometimes $>1,000$ times faster than standard simulation \cite{Ding2010}. 
In a similar spirit, tabulation of a scaled form of the generalized Born implicit-solvent model was shown to lead to significant speed gain for the tabulated part \cite{Larsson2010}.


\section{Concluding Perspective}

The main goal of this review has been to array various methodologies into qualitative groupings so as to aid the critical analysis necessary to make progress in equilibrium sampling of biomolecules. 
Where appropriate, an effort has been made to offer a point of view on essential strengths and weaknesses of various methods.
As an example, when considering popular multi-level schemes (Secs.\ \ref{sec:repex} - \ref{sec:resex}), users should be confident that some level of the ``ladder'' can be well sampled.

Interesting conclusions result from surveying the sampling literature.
Except in small systems, purely algorithmic improvements have yet to demonstrably accelerate equilibrium sampling of biomolecules by a significant amount. 
Hardware-based advances have been more dramatic, however.
In fairness, demonstrating the effectiveness of new hardware for MD is much more straightforward than assessing an algorithm.

To aid future progress, developing and using sampling ``yardsticks'' should be a key priority for the field (Sec.\ \ref{sec:ess}).
Such measures should probe the configuration-space distribution in an objective, automatic way to measure the effective sample size.
Once a small number of standard measures of sampling quality are accepted \underline{and used}, efforts naturally will focus on approaches that make a significant difference.
Currently, there is a proliferation of nuanced modifications of a small number of central ideas, without a good basis for distinguishing among them.
Unlike in the history of theoretical physics, where elegance has sometimes served as a guide for truth, the sampling problem cries out for a ``bottom line'' focus on efficiency.
After all, it is this efficiency which ultimately will permit addressing biophysical and biochemical questions with confidence.

In summary, there seems little choice but to express pessimism on the current state of equilibrium sampling of important biomolecular systems.
Keys to moving forward would seem to be (i) exploiting hardware; (ii) quantifying sampling to determine which algorithms are more efficient than MD; and (iii) employing large-resource simulation data to provide benchmarks and guide future efforts.

\section*{Acknowledgments}
The author has benefitted from numerous stimulating discussions of sampling with colleagues and group members over the years.
Financial support has been provided by grants from the National Science Foundation (Grant No.\ MCB-0643456) and the National Institutes of Health (Grant No.\ GM076569).

\newpage
\bibliographystyle{/usr/share/texmf/bibtex/bst/base/abbrv.bst}
\bibliography{../../bib/dmz}

\begin{thebibliography}{100}

\bibitem{Abrams2008}
J.~B. Abrams and M.~E. Tuckerman.
\newblock Efficient and direct generation of multidimensional free energy
  surfaces via adiabatic dynamics without coordinate transformations.
\newblock {\em The Journal of Physical Chemistry B}, 112(49):15742--15757, Dec.
  2008.

\bibitem{Allen-Tildesley}
M.~P. Allen and D.~J. Tildesley.
\newblock {\em Computer Simulation of Liquids}.
\newblock Oxford University Press, Oxford, 1987.

\bibitem{Beck2007}
D.~A.~C. Beck, G.~W.~N. White, and V.~Daggett.
\newblock Exploring the energy landscape of protein folding using
  replica-exchange and conventional molecular dynamics simulations.
\newblock {\em J Struct Biol}, 157(3):514--523, Mar 2007.

\bibitem{Straub-1997b}
B.~J. Berne and J.~E. Straub.
\newblock Novel methods of sampling phase space in the simulation of biological
  systems.
\newblock {\em Curr Opin Struc Biol}, 7:181--189, 1997.

\bibitem{Betancourt2005}
M.~R. Betancourt.
\newblock Efficient {Monte Carlo} trial moves for polypeptide simulations.
\newblock {\em J Chem Phys}, 123(17):174905, Nov 2005.

\bibitem{Bhatt2010a}
D.~Bhatt, B.~W. Zhang, and D.~M. Zuckerman.
\newblock Steady state via weighted ensemble path sampling.
\newblock {\em Journal of Chemical Physics}, 133:014110, 2010.

\bibitem{Bogdan2006}
T.~V. Bogdan, D.~J. Wales, and F.~Calvo.
\newblock Equilibrium thermodynamics from basin-sampling.
\newblock {\em J Chem Phys}, 124(4):044102, Jan 2006.

\bibitem{Brenner2007}
P.~Brenner, C.~R. Sweet, D.~VonHandorf, and J.~A. Izaguirre.
\newblock Accelerating the replica exchange method through an efficient
  all-pairs exchange.
\newblock {\em J Chem Phys}, 126(7):074103, Feb 2007.

\bibitem{Head-Gordon-2002}
S.~Brown and T.~Head-Gordon.
\newblock Cool walking: A new {Markov} chain {{Monte Carlo}} sampling method.
\newblock {\em J. Comp. Chem.}, 24:68--76, 2002.

\bibitem{Bussi2006a}
G.~Bussi, A.~Laio, and M.~Parrinello.
\newblock Equilibrium free energies from nonequilibrium metadynamics.
\newblock {\em Phys Rev Lett}, 96(9):090601, Mar 2006.

\bibitem{Calvo2005}
F.~Calvo.
\newblock All-exchanges parallel tempering.
\newblock {\em J Chem Phys}, 123(12):124106, Sep 2005.

\bibitem{Meirovitch-2004b}
S.~Cheluvaraja and H.~Meirovitch.
\newblock Simulation method for calculating the entropy and free energy
  peptides and proteins.
\newblock {\em Proc. Nat. Acad. Sci.}, 101:9241--9246, 2004.

\bibitem{Chodera2007}
J.~D. Chodera, N.~Singhal, V.~S. Pande, K.~A. Dill, and W.~C. Swope.
\newblock Automatic discovery of metastable states for the construction of
  markov models of macromolecular conformational dynamics.
\newblock {\em The Journal of Chemical Physics}, 126(15):155101--17, 2007.

\bibitem{Chodera2007a}
J.~D. Chodera, W.~C. Swope, J.~W. Pitera, C.~Seok, and K.~A. Dill.
\newblock Use of the weighted histogram analysis method for the analysis of
  simulated and parallel tempering simulations.
\newblock {\em Journal of Chemical Theory and Computation}, 3(1):26--41, 2007.

\bibitem{Christen2006}
M.~Christen and W.~F. van Gunsteren.
\newblock Multigraining: an algorithm for simultaneous fine-grained and
  coarse-grained simulation of molecular systems.
\newblock {\em J Chem Phys}, 124(15):154106, Apr 2006.

\bibitem{Christen2008}
M.~Christen and W.~F. van Gunsteren.
\newblock On searching in, sampling of, and dynamically moving through
  conformational space of biomolecular systems: A review.
\newblock {\em J. Comput. Chem.}, 29(2):157--166, 2008.

\bibitem{Das2008}
R.~Das and D.~Baker.
\newblock Macromolecular modeling with {Rosetta}.
\newblock {\em Annu Rev Biochem}, 77:363--382, 2008.

\bibitem{Denschlag2008}
R.~Denschlag, M.~Lingenheil, and P.~Tavan.
\newblock Efficiency reduction and pseudo-convergence in replica exchange
  sampling of peptide folding-unfolding equilibria.
\newblock {\em Chemical Physics Letters}, 458:244--248, 2008.

\bibitem{Ding2010}
Y.~Ding, A.~B. Mamonov, and D.~M. Zuckerman.
\newblock Efficient equilibrium sampling of all-atom peptides using
  library-based {{Monte Carlo}}.
\newblock {\em J Phys Chem B}, 114(17):5870--5877, May 2010.
\newblock {PMCID: PMC2882875}.

\bibitem{Dror2009a}
R.~O. Dror, D.~H. Arlow, D.~W. Borhani, M.~Ã. Jensen, S.~Piana, and D.~E. Shaw.
\newblock Identification of two distinct inactive conformations of the
  beta2-adrenergic receptor reconciles structural and biochemical observations.
\newblock {\em Proc Natl Acad Sci U S A}, 106(12):4689--4694, Mar 2009.

\bibitem{Duan1998}
Y.~Duan and P.~A. Kollman.
\newblock Pathways to a protein folding intermediate observed in a
  1-microsecond simulation in aqueous solution.
\newblock {\em Science}, 282(5389):740--744, Oct 1998.

\bibitem{Earl2008}
D.~J. Earl and M.~W. Deem.
\newblock {Monte Carlo} simulations.
\newblock {\em Methods Mol Biol}, 443:25--36, 2008.

\bibitem{AChang2007}
C.~en~A~Chang, W.~Chen, and M.~K. Gilson.
\newblock Ligand configurational entropy and protein binding.
\newblock {\em Proc Natl Acad Sci U S A}, 104(5):1534--1539, Jan 2007.

\bibitem{Ensign2007}
D.~L. Ensign, P.~M. Kasson, and V.~S. Pande.
\newblock Heterogeneity even at the speed limit of folding: large-scale
  molecular dynamics study of a fast-folding variant of the villin headpiece.
\newblock {\em J Mol Biol}, 374(3):806--816, Nov 2007.

\bibitem{Swendsen-2004}
M.~Fasnacht, R.~H. Swendsen, and J.~M. Rosenberg.
\newblock Adaptive integration method for {M}onte {C}arlo simulations.
\newblock {\em Phys. Rev. E}, 69:056704, 2004.

\bibitem{Fenwick2003}
M.~K. Fenwick and F.~A. Escobedo.
\newblock Expanded ensemble and replica exchange methods for simulation of
  protein-like systems.
\newblock {\em The Journal of Chemical Physics}, 119(22):11998--12010, 2003.

\bibitem{Swendsen-1988}
A.~M. Ferrenberg and R.~H. Swendsen.
\newblock New {M}onte {C}arlo technique for studying phase transitions.
\newblock {\em Phys. Rev. Lett.}, 61:2635--2638, 1988.

\bibitem{Doll-1990}
D.~D. Frantz, D.~L. Freeman, and J.~D. Doll.
\newblock Reducing quasi-ergodic behavior in {{Monte Carlo}} simulation by
  {J}-walking: applications to atomic clusters.
\newblock {\em J. Chem. Phys.}, 93:2769--2784, 1990.

\bibitem{Frauenkron1998}
H.~Frauenkron, U.~Bastolla, E.~Gerstner, P.~Grassberger, and W.~Nadler.
\newblock New {{Monte Carlo}} algorithm for protein folding.
\newblock {\em Physical Review Letters}, 80(14):3149--3152, 1998.

\bibitem{Freddolino2008}
P.~L. Freddolino, F.~Liu, M.~Gruebele, and K.~Schulten.
\newblock Ten-microsecond molecular dynamics simulation of a fast-folding ww
  domain.
\newblock {\em Biophys J}, 94(10):L75--L77, May 2008.

\bibitem{Freddolino2009}
P.~L. Freddolino, S.~Park, B.~Roux, and K.~Schulten.
\newblock Force field bias in protein folding simulations.
\newblock {\em Biophys J}, 96(9):3772--3780, May 2009.

\bibitem{Frenkel-book}
D.~Frenkel and B.~Smit.
\newblock {\em Understanding Molecular Simulation}.
\newblock Academic Press, San Diego, 1996.

\bibitem{Friedrichs2009}
M.~S. Friedrichs, P.~Eastman, V.~Vaidyanathan, M.~Houston, S.~Legrand, A.~L.
  Beberg, D.~L. Ensign, C.~M. Bruns, and V.~S. Pande.
\newblock Accelerating molecular dynamic simulation on graphics processing
  units.
\newblock {\em J Comput Chem}, 2009.

\bibitem{Geyer1991}
C.~J. Geyer.
\newblock Markov chain {{Monte Carlo}} maximum likelihood.
\newblock In E.~M. Keramidas, editor, {\em Proceedings of the 23rd symposium on
  the interface}, Computing science and statistics, 1991.

\bibitem{Grassberger1997}
P.~Grassberger.
\newblock Pruned-enriched rosenbluth method: Simulations of theta polymers of
  chain length up to 1 000 000.
\newblock {\em Physical Review E (Statistical Physics, Plasmas, Fluids, and
  Related Interdisciplinary Topics)}, 56(3):3682--3693, 1997.

\bibitem{Grossfield2007a}
A.~Grossfield, S.~E. Feller, and M.~C. Pitman.
\newblock Convergence of molecular dynamics simulations of membrane proteins.
\newblock {\em Proteins}, 67(1):31--40, Apr 2007.

\bibitem{Grossfield2008}
A.~Grossfield, M.~C. Pitman, S.~E. Feller, O.~Soubias, and K.~Gawrisch.
\newblock Internal hydration increases during activation of the
  {G}-protein-coupled receptor rhodopsin.
\newblock {\em J Mol Biol}, 381(2):478--486, Aug 2008.

\bibitem{Grossfield2009}
A.~Grossfield and D.~M. Zuckerman.
\newblock Quantifying uncertainty and sampling quality in biomolecular
  simulations.
\newblock {\em Annu Rep Comput Chem}, 5:23--48, Jan 2009.
\newblock {PMCID: PMC2865156}.

\bibitem{Hamelberg2004a}
D.~Hamelberg, J.~Mongan, and J.~A. McCammon.
\newblock Accelerated molecular dynamics: a promising and efficient simulation
  method for biomolecules.
\newblock {\em J Chem Phys}, 120(24):11919--11929, Jun 2004.

\bibitem{Hansmann1996}
Hansmann and Okamoto.
\newblock Monte carlo simulations in generalized ensemble: Multicanonical
  algorithm versus simulated tempering.
\newblock {\em Phys Rev E Stat Phys Plasmas Fluids Relat Interdiscip Topics},
  54(5):5863--5865, Nov 1996.

\bibitem{Hess2002}
B.~Hess.
\newblock Convergence of sampling in protein simulations.
\newblock {\em Physical Review E}, 65(3):031910, 2002.

\bibitem{Hinrichs2007}
N.~S. Hinrichs and V.~S. Pande.
\newblock Calculation of the distribution of eigenvalues and eigenvectors in
  markovian state models for molecular dynamics.
\newblock {\em J Chem Phys}, 126(24):244101, Jun 2007.

\bibitem{Hnizdo2008}
V.~Hnizdo, J.~Tan, B.~J. Killian, and M.~K. Gilson.
\newblock Efficient calculation of configurational entropy from molecular
  simulations by combining the mutual-information expansion and
  nearest-neighbor methods.
\newblock {\em J Comput Chem}, 29(10):1605--1614, Jul 2008.

\bibitem{Howard2001}
J.~Howard.
\newblock {\em Mechanics of Motor Proteins and the Cytoskeleton}.
\newblock Sinauer Associate, Sunderland, Mass, 2001.

\bibitem{Hritz2008a}
J.~Hritz and C.~Oostenbrink.
\newblock Hamiltonian replica exchange molecular dynamics using soft-core
  interactions.
\newblock {\em J Chem Phys}, 128(14):144121, Apr 2008.

\bibitem{Hu2006}
J.~Hu, A.~Ma, and A.~R. Dinner.
\newblock {Monte Carlo} simulations of biomolecules: The {MC} module in
  {CHARMM}.
\newblock {\em J. Comput. Chem.}, 27(2):203--216, 2006.

\bibitem{Huang2009}
X.~Huang, G.~R. Bowman, S.~Bacallado, and V.~S. Pande.
\newblock Rapid equilibrium sampling initiated from nonequilibrium data.
\newblock {\em Proc Natl Acad Sci U S A}, 106(47):19765--19769, Nov 2009.

\bibitem{Huang2008}
X.~Huang, G.~R. Bowman, and V.~S. Pande.
\newblock Convergence of folding free energy landscapes via application of
  enhanced sampling methods in a distributed computing environment.
\newblock {\em J Chem Phys}, 128(20):205106, May 2008.

\bibitem{Huber1997}
G.~A. Huber and J.~A. McCammon.
\newblock Weighted-ensemble simulated annealing: Faster optimization on
  hierarchical energy surfaces.
\newblock {\em Physical Review E}, 55(4):4822, 1997.

\bibitem{Huber1994}
T.~Huber, A.~E. Torda, and W.~F. van Gunsteren.
\newblock Local elevation: a method for improving the searching properties of
  molecular dynamics simulation.
\newblock {\em J Comput Aided Mol Des}, 8(6):695--708, Dec 1994.

\bibitem{Nemoto-1996}
K.~Hukushima and K.~Nemoto.
\newblock Exchange {M}onte {C}arlo method and application to spin glass
  simulation.
\newblock {\em J.\ Phys.\ Soc.\ Jpn.}, 65:1604--1608, 1996.

\bibitem{Jarzynski-1997a}
C.~Jarzynski.
\newblock Nonequilibrium equality for free energy differences.
\newblock {\em Phys. Rev. Lett.}, 78:2690--2693, 1997.

\bibitem{Jorgensen-1996b}
W.~L. Jorgensen and J.~Tirado-Rives.
\newblock {{Monte Carlo}} vs molecular dynamics for conformational sampling.
\newblock {\em J. Phys. Chem.}, 100:14508--14513, 1996.

\bibitem{Killian2007}
B.~J. Killian, J.~Y. Kravitz, and M.~K. Gilson.
\newblock Extraction of configurational entropy from molecular simulations via
  an expansion approximation.
\newblock {\em J Chem Phys}, 127(2):024107, Jul 2007.

\bibitem{Kim2006}
J.~Kim, J.~E. Straub, and T.~Keyes.
\newblock Statistical-temperature {Monte Carlo} and molecular dynamics
  algorithms.
\newblock {\em Phys Rev Lett}, 97(5):050601, Aug 2006.

\bibitem{Oshiro-1997}
R.~M.~A. Knegtel, I.~D. Kuntz, and C.~M. Oshiro.
\newblock Molecular docking to ensembles of protein structures.
\newblock {\em J. Mol. Bio.}, 266:424--440, 1997.

\bibitem{Kofke2002}
D.~A. Kofke.
\newblock On the acceptance probability of replica-exchange monte carlo trials.
\newblock {\em The Journal of Chemical Physics}, 117(15):6911--6914, 2002.

\bibitem{Brooks-1996}
X.~Kong and C.~M. Brooks.
\newblock Lambda-dynamics: A new approach to free energy calculations.
\newblock {\em J. Chem. Phys.}, 105:2414--2423, 1996.

\bibitem{Swendsen-1992}
S.~Kumar, D.~Bouzida, R.~H. Swendsen, P.~A. Kollman, and J.~M. Rosenberg.
\newblock The weighted histogram analysis method for free-energy calculations
  on biomolecules. i. the method.
\newblock {\em J. Comp. Chem.}, 13:1011--1021, 1992.

\bibitem{Laio2002}
A.~Laio and M.~Parrinello.
\newblock Escaping free-energy minima.
\newblock {\em Proc Natl Acad Sci U S A}, 99(20):12562--12566, Oct 2002.

\bibitem{Larsson2010}
P.~Larsson and E.~Lindahl.
\newblock A high-performance parallel-generalized born implementation enabled
  by tabulated interaction rescaling.
\newblock {\em J. Comput. Chem.}, 31(14):2593--2600, 2010.

\bibitem{Leach2001}
A.~R. Leach.
\newblock {\em Molecular Modelling: Principles and Applications}.
\newblock Prentice Hall, 2001.

\bibitem{Leimkuhler2004}
B.~J. Leimkuhler and C.~R. Sweet.
\newblock The canonical ensemble via symplectic integrators using nose and
  nose-poincare chains.
\newblock {\em The Journal of Chemical Physics}, 121(1):108--116, 2004.

\bibitem{Li2006b}
H.~Li, G.~Li, B.~A. Berg, and W.~Yang.
\newblock Finite reservoir replica exchange to enhance canonical sampling in
  rugged energy surfaces.
\newblock {\em J Chem Phys}, 125(14):144902, Oct 2006.

\bibitem{Lin2002}
J.-H. Lin, A.~L. Perryman, J.~R. Schames, and J.~A. McCammon.
\newblock Computational drug design accommodating receptor flexibility: the
  relaxed complex scheme.
\newblock {\em J Am Chem Soc}, 124(20):5632--5633, May 2002.

\bibitem{Liu2002}
J.~S. Liu.
\newblock {\em {{Monte Carlo}} Strategies in Scientific Computing}.
\newblock Springer, New York, 2002.

\bibitem{Berne-2005}
P.~Liu, B.~Kim, R.~A. Friesner, and B.~J. Berne.
\newblock Replica exchange with solute tempering: A method for sampling
  biological systems in explicit water.
\newblock {\em Proc. Nat. Acad. Sci.}, 102:13749--13754, 2005.

\bibitem{Liu2008b}
P.~Liu, Q.~Shi, E.~Lyman, and G.~A. Voth.
\newblock Reconstructing atomistic detail for coarse-grained models with
  resolution exchange.
\newblock {\em J Chem Phys}, 129(11):114103, Sep 2008.

\bibitem{Liu1993}
Z.~Liu and B.~J. Berne.
\newblock Method for accelerating chain folding and mixing.
\newblock {\em The Journal of Chemical Physics}, 99(8):6071--6077, 1993.

\bibitem{Zuckerman-2006a}
E.~Lyman, F.~M. Ytreberg, and D.~M. Zuckerman.
\newblock Resolution exchange simulation.
\newblock {\em Phys. Rev. Lett.}, 96:028105, 2006.

\bibitem{Zuckerman-2006c}
E.~Lyman and D.~M. Zuckerman.
\newblock Resolution exchange simulation with incremental coarsening.
\newblock {\em J. Chem. Theory Comp.}, 2:656--666, 2006.

\bibitem{Lyman2007}
E.~Lyman and D.~M. Zuckerman.
\newblock Annealed importance sampling of peptides.
\newblock {\em The Journal of Chemical Physics}, 127(6):065101--6, 2007.

\bibitem{Lyman2007a}
E.~Lyman and D.~M. Zuckerman.
\newblock On the structural convergence of biomolecular simulations by
  determination of the effective sample size.
\newblock {\em J. Phys. Chem. B}, 111(44):12876--12882, 2007.
\newblock {PMCID: PMC2538559}.

\bibitem{Lyman2009}
E.~Lyman and D.~M. Zuckerman.
\newblock Resampling improves the efficiency of a ``fast-switch'' equilibrium
  sampling protocol.
\newblock {\em J Chem Phys}, 130(8):081102, Feb 2009.
\newblock {PMCID: PMC2671214}.

\bibitem{Lyubartsev1992}
A.~P. Lyubartsev, A.~A. Martsinovski, S.~V. Shevkunov, and P.~N.
  Vorontsov-Velyaminov.
\newblock New approach to {Monte Carlo} calculation of the free energy:
  {Method} of expanded ensembles.
\newblock {\em The Journal of Chemical Physics}, 96(3):1776--1783, 1992.

\bibitem{Machta2009}
J.~Machta.
\newblock Strengths and weaknesses of parallel tempering.
\newblock {\em Phys Rev E Stat Nonlin Soft Matter Phys}, 80(5 Pt 2):056706, Nov
  2009.

\bibitem{Mamonov2009b}
A.~B. Mamonov, D.~Bhatt, D.~J. Cashman, Y.~Ding, and D.~M. Zuckerman.
\newblock General library-based {{Monte Carlo}} technique enables equilibrium
  sampling of semi-atomistic protein models.
\newblock {\em J Phys Chem B}, 113(31):10891--10904, Aug 2009.
\newblock {PMCID: PMC2766542}.

\bibitem{Mamonov2009a}
A.~B. Mamonov, X.~Zhang, and D.~Zuckerman.
\newblock Rapid sampling of all-atom peptides using a library-based
  polymer-growth approach.
\newblock {\em Journal of Computational Chemistry}, Accepted., 2010.
\newblock {In press. P}reprint available: http://arxiv.org/abs/0910.2495.
  {PMCID in process}.

\bibitem{Manousiouthakis1999}
V.~I. Manousiouthakis and M.~W. Deem.
\newblock Strict detailed balance is unnecessary in {Monte Carlo} simulation.
\newblock {\em The Journal of Chemical Physics}, 110(6):2753--2756, 1999.

\bibitem{Mao2010}
A.~H. Mao, S.~L. Crick, A.~Vitalis, C.~L. Chicoine, and R.~V. Pappu.
\newblock Net charge per residue modulates conformational ensembles of
  intrinsically disordered proteins.
\newblock {\em Proc Natl Acad Sci U S A}, 107(18):8183--8188, May 2010.

\bibitem{Maragliano2008}
L.~Maragliano and E.~Vanden-Eijnden.
\newblock Single-sweep methods for free energy calculations.
\newblock {\em J Chem Phys}, 128(18):184110, May 2008.

\bibitem{McCammon1977}
J.~A. McCammon, B.~R. Gelin, and M.~Karplus.
\newblock Dynamics of folded proteins.
\newblock {\em Nature}, 267(5612):585--590, 1977.

\bibitem{Meirovitch1982}
H.~Meirovitch.
\newblock A new method for simulation of real chains. scanning future steps.
\newblock {\em J Phys A}, 15:L735--L740, 1982.

\bibitem{Minary2004a}
P.~Minary, M.~E. Tuckerman, and G.~J. Martyna.
\newblock Long time molecular dynamics for enhanced conformational sampling in
  biomolecular systems.
\newblock {\em Phys Rev Lett}, 93(15):150201, Oct 2004.

\bibitem{Mitsutake2001}
A.~Mitsutake, Y.~Sugita, and Y.~Okamoto.
\newblock Generalized-ensemble algorithms for molecular simulations of
  biopolymers.
\newblock {\em Biopolymers}, 60(2):96--123, 2001.

\bibitem{Mountain1989}
R.~D. Mountain and D.~Thirumalai.
\newblock Measures of effective ergodic convergence in liquids.
\newblock {\em J. Phys. Chem.}, 93(19):6975--6979, 1989.

\bibitem{Neal-2001}
R.~M. Neal.
\newblock Annealed importance sampling.
\newblock {\em Statistics and Computing}, 11:125--139, 2001.

\bibitem{Neirotti2000}
J.~P. Neirotti, D.~L. Freeman, and J.~D. Doll.
\newblock Approach to ergodicity in {{Monte Carlo}} simulations.
\newblock {\em Physical Review E}, 62(5):7445, 2000.

\bibitem{Noe2009}
F.~Noe, C.~Schütte, E.~Vanden-Eijnden, L.~Reich, and T.~R. Weikl.
\newblock Constructing the equilibrium ensemble of folding pathways from short
  off-equilibrium simulations.
\newblock {\em Proc Natl Acad Sci U S A}, 106(45):19011--19016, Nov 2009.

\bibitem{Nowak2000}
Nowak, Chantrell, and Kennedy.
\newblock {Monte Carlo} simulation with time step quantification in terms of
  {Langevin} dynamics.
\newblock {\em Phys Rev Lett}, 84(1):163--166, Jan 2000.

\bibitem{Noe2007}
F.~Noé, I.~Horenko, C.~Schütte, and J.~C. Smith.
\newblock Hierarchical analysis of conformational dynamics in biomolecules:
  transition networks of metastable states.
\newblock {\em J Chem Phys}, 126(15):155102, Apr 2007.

\bibitem{Nymeyer2008}
H.~Nymeyer.
\newblock How efficient is replica exchange molecular dynamics? an analytic
  approach.
\newblock {\em J. Chem. Theory Comput.}, 4(4):626--636, 2008.

\bibitem{Okur2007}
A.~Okur, D.~R. Roe, G.~Cui, V.~Hornak, and C.~Simmerling.
\newblock Improving convergence of replica-exchange simulations through
  coupling to a high-temperature structure reservoir.
\newblock {\em Journal of Chemical Theory and Computation}, 3(2):557--568,
  2007.

\bibitem{Opps2001}
S.~B. Opps and J.~Schofield.
\newblock Extended state-space monte carlo methods.
\newblock {\em Phys Rev E Stat Nonlin Soft Matter Phys}, 63(5 Pt 2):056701, May
  2001.

\bibitem{Garcia-2004}
D.~Paschek and A.~E. Garcia.
\newblock Reversible temperature and pressure denaturation of a protein
  fragment: A replica-exchange molecular dynamics simulation study.
\newblock {\em Phys. Rev. Lett.}, 93:238105, 2004.

\bibitem{Periole2007a}
X.~Periole and A.~E. Mark.
\newblock Convergence and sampling efficiency in replica exchange simulations
  of peptide folding in explicit solvent.
\newblock {\em J Chem Phys}, 126(1):014903, Jan 2007.

\bibitem{Rahman2002}
J.~A. Rahman and J.~C. Tully.
\newblock Puddle-skimming: An efficient sampling of multidimensional
  configuration space.
\newblock {\em The Journal of Chemical Physics}, 116(20):8750--8760, 2002.

\bibitem{Raiteri2006}
P.~Raiteri, A.~Laio, F.~L. Gervasio, C.~Micheletti, and M.~Parrinello.
\newblock Efficient reconstruction of complex free energy landscapes by
  multiple walkers metadynamics.
\newblock {\em J Phys Chem B}, 110(8):3533--3539, Mar 2006.

\bibitem{Rathore2005}
N.~Rathore, M.~Chopra, and J.~J. de~Pablo.
\newblock Optimal allocation of replicas in parallel tempering simulations.
\newblock {\em J Chem Phys}, 122(2):024111, Jan 2005.

\bibitem{Rhee2003}
Y.~M. Rhee and V.~S. Pande.
\newblock Multiplexed-replica exchange molecular dynamics method for protein
  folding simulation.
\newblock {\em Biophys J}, 84(2 Pt 1):775--786, Feb 2003.

\bibitem{Rick2007}
S.~W. Rick.
\newblock Replica exchange with dynamical scaling.
\newblock {\em J Chem Phys}, 126(5):054102, Feb 2007.

\bibitem{Rosta2009b}
E.~Rosta and G.~Hummer.
\newblock Error and efficiency of replica exchange molecular dynamics
  simulations.
\newblock {\em J Chem Phys}, 131(16):165102, Oct 2009.

\bibitem{Ruscio2010}
J.~Z. Ruscio, N.~L. Fawzi, and T.~Head-Gordon.
\newblock How hot? systematic convergence of the replica exchange method using
  multiple reservoirs.
\newblock {\em J. Comput. Chem.}, 31(3):620--627, 2010.

\bibitem{Sanbonmatsu2002}
K.~Y. Sanbonmatsu and A.~E. García.
\newblock Structure of met-enkephalin in explicit aqueous solution using
  replica exchange molecular dynamics.
\newblock {\em Proteins}, 46(2):225--234, Feb 2002.

\bibitem{Schlick2002}
T.~Schlick.
\newblock {\em Molecular Modeling and Simulation}.
\newblock Springer, 2002.

\bibitem{Schlick1998}
T.~Schlick, M.~Mandziuk, R.~D. Skeel, and K.~Srinivas.
\newblock Nonlinear resonance artifacts in molecular dynamics simulations.
\newblock {\em Journal of Computational Physics}, 140(1):1--29, Feb. 1998.

\bibitem{Shaw2007}
D.~E. Shaw, M.~M. Deneroff, R.~O. Dror, J.~S. Kuskin, R.~H. Larson, J.~K.
  Salmon, C.~Young, B.~Batson, K.~J. Bowers, J.~C. Chao, M.~P. Eastwood,
  J.~Gagliardo, J.~P. Grossman, C.~R. Ho, D.~J. Ierardi, I.~Kolossv\'{a}ry,
  J.~L. Klepeis, T.~Layman, C.~McLeavey, M.~A. Moraes, R.~Mueller, E.~C.
  Priest, Y.~Shan, J.~Spengler, M.~Theobald, B.~Towles, and S.~C. Wang.
\newblock Anton, a special-purpose machine for molecular dynamics simulation.
\newblock In {\em ISCA '07: Proceedings of the 34th annual international
  symposium on Computer architecture}, pages 1--12, New York, NY, USA, 2007.
  ACM.

\bibitem{Shaw2009}
D.~E. Shaw, R.~O. Dror, J.~K. Salmon, J.~P. Grossman, K.~M. Mackenzie, J.~A.
  Bank, C.~Young, M.~M. Deneroff, B.~Batson, K.~J. Bowers, E.~Chow, M.~P.
  Eastwood, D.~J. Ierardi, J.~L. Klepeis, J.~S. Kuskin, R.~H. Larson,
  K.~Lindorff-Larsen, P.~Maragakis, M.~A. Moraes, S.~Piana, Y.~Shan, and
  B.~Towles.
\newblock Millisecond-scale molecular dynamics simulations on anton.
\newblock In {\em SC '09: Proceedings of the Conference on High Performance
  Computing Networking, Storage and Analysis}, pages 1--11, New York, NY, USA,
  2009. ACM.

\bibitem{Shen2008}
T.~Shen and D.~Hamelberg.
\newblock A statistical analysis of the precision of reweighting-based
  simulations.
\newblock {\em J Chem Phys}, 129(3):034103, Jul 2008.

\bibitem{Shimada2001}
J.~Shimada, E.~L. Kussell, and E.~I. Shakhnovich.
\newblock The folding thermodynamics and kinetics of crambin using an all-atom
  {{Monte Carlo}} simulation.
\newblock {\em J Mol Biol}, 308(1):79--95, 2001.

\bibitem{Shirts2008}
M.~R. Shirts and J.~D. Chodera.
\newblock Statistically optimal analysis of samples from multiple equilibrium
  states.
\newblock {\em J Chem Phys}, 129(12):124105, Sep 2008.

\bibitem{Sindhikara2008}
D.~Sindhikara, Y.~Meng, and A.~E. Roitberg.
\newblock Exchange frequency in replica exchange molecular dynamics.
\newblock {\em J Chem Phys}, 128(2):024103, Jan 2008.

\bibitem{Smith2008}
C.~A. Smith and T.~Kortemme.
\newblock Backrub-like backbone simulation recapitulates natural protein
  conformational variability and improves mutant side-chain prediction.
\newblock {\em J Mol Biol}, 380(4):742--756, Jul 2008.

\bibitem{vanGunsteren-2002}
L.~J. Smith, X.~Daura, and W.~F. van Gunsteren.
\newblock Assessing equilibration and convergence in biomolecular simulations.
\newblock {\em Proteins}, 48:487--496, 2002.

\bibitem{Snow2002a}
C.~D. Snow, H.~Nguyen, V.~S. Pande, and M.~Gruebele.
\newblock Absolute comparison of simulated and experimental protein-folding
  dynamics.
\newblock {\em Nature}, 420(6911):102--106, Nov 2002.

\bibitem{Straub1993}
J.~E. Straub and D.~Thirumalai.
\newblock Exploring the energy landscape in proteins.
\newblock {\em Proceedings of the National Academy of Sciences},
  90(3):809--813, 1993.

\bibitem{Okamoto-2000}
Y.~Sugita, A.~Kitao, and Y.~Okamoto.
\newblock Multidimensional replica-exchange method for free-energy
  calculations.
\newblock {\em J. Chem. Phys.}, 113:6042--6051, 2000.

\bibitem{Swendsen-1986}
R.~H. Swendsen and J.-S. Wang.
\newblock Replica {M}onte {C}arlo simulation of spin-glasses.
\newblock {\em Phys. Rev. Lett.}, 57:2607--2609, 1986.

\bibitem{Teleman1986}
O.~Teleman and B.~Jonsson.
\newblock Vectorizing a general purpose molecular dynamics simulation program.
\newblock {\em J. Comput. Chem.}, 7(1):58--66, 1986.

\bibitem{Tidor-1993}
B.~Tidor.
\newblock Simulated annealing on free energy surfaces by a combined molecular
  dynamics and {{Monte Carlo}} approach.
\newblock {\em J. Phys. Chem.}, 97:1069--173, 1993.

\bibitem{Trebst2006}
S.~Trebst, M.~Troyer, and U.~H.~E. Hansmann.
\newblock Optimized parallel tempering simulations of proteins.
\newblock {\em J Chem Phys}, 124(17):174903, May 2006.

\bibitem{Tuckerman1992}
M.~Tuckerman, B.~J. Berne, and G.~J. Martyna.
\newblock Reversible multiple time scale molecular dynamics.
\newblock {\em The Journal of Chemical Physics}, 97(3):1990--2001, 1992.

\bibitem{Ulmschneider2003}
J.~P. Ulmschneider and W.~L. Jorgensen.
\newblock {Monte Carlo} backbone sampling for polypeptides with variable bond
  angles and dihedral angles using concerted rotations and a gaussian bias.
\newblock {\em The Journal of Chemical Physics}, 118(9):4261--4271, 2003.

\bibitem{Ulmschneider2006}
J.~P. Ulmschneider, M.~B. Ulmschneider, and A.~D. Nola.
\newblock {Monte Carlo} vs molecular dynamics for all-atom polypeptide folding
  simulations.
\newblock {\em J Phys Chem B}, 110(33):16733--16742, Aug 2006.

\bibitem{Valeriani2007}
C.~Valeriani, R.~J. Allen, M.~J. Morelli, D.~Frenkel, and P.~R. ten Wolde.
\newblock Computing stationary distributions in equilibrium and nonequilibrium
  systems with forward flux sampling.
\newblock {\em J Chem Phys}, 127(11):114109, Sep 2007.

\bibitem{Velikson1992}
B.~Velikson, T.~Garel, J.~C. Niel, H.~Orland, and J.~C. Smith.
\newblock Conformational distribution of heptaalanine: Analysis using a new
  {{Monte Carlo}} chain growth method.
\newblock {\em Journal of Computational Chemistry}, 13(10):1216--1233, 1992.

\bibitem{Vendruscolo1997}
M.~Vendruscolo.
\newblock Modified configurational bias {Monte Carlo} method for simulation of
  polymer systems.
\newblock {\em The Journal of Chemical Physics}, 106(7):2970--2976, 1997.

\bibitem{Vitalis2009}
A.~Vitalis and R.~V. Pappu.
\newblock Methods for {Monte Carlo} simulations of biomacromolecules.
\newblock {\em Annu Rep Comput Chem}, 5:49--76, Jan 2009.

\bibitem{Voter1997}
A.~F. Voter.
\newblock Hyperdynamics: Accelerated molecular dynamics of infrequent events.
\newblock {\em Phys. Rev. Lett.}, 78(20):3908--3911, May 1997.

\bibitem{Wales2004}
D.~Wales.
\newblock {\em Energy Landscapes: Applications to Clusters, Biomolecules and
  Glasses}.
\newblock Cambridge University Press, 2004.

\bibitem{Wall1959}
F.~T. Wall and J.~J. Erpenbeck.
\newblock New method for the statistical computation of polymer dimensions.
\newblock {\em The Journal of Chemical Physics}, 30(3):634--637, 1959.

\bibitem{Wang2001}
F.~Wang and D.~P. Landau.
\newblock Efficient, multiple-range random walk algorithm to calculate the
  density of states.
\newblock {\em Phys. Rev. Lett.}, 86(10):2050--, Mar. 2001.

\bibitem{Warmflash2007}
A.~Warmflash, P.~Bhimalapuram, and A.~R. Dinner.
\newblock Umbrella sampling for nonequilibrium processes.
\newblock {\em The Journal of Chemical Physics}, 127(15):154112--8, 2007.

\bibitem{Wu1999}
M.~G. Wu and M.~W. Deem.
\newblock Analytical rebridging {Monte Carlo}: Application to cis/trans
  isomerization in proline-containing, cyclic peptides.
\newblock {\em The Journal of Chemical Physics}, 111(14):6625--6632, 1999.

\bibitem{Ytreberg2007}
F.~M. Ytreberg, S.~K. Aroutiounian, and D.~M. Zuckerman.
\newblock Demonstrated convergence of the equilibrium ensemble for a fast
  united-residue protein model.
\newblock {\em J. Chem. Theory Comput.}, 3(5):1860--1866, 2007.

\bibitem{Ytreberg2008}
F.~M. Ytreberg and D.~M. Zuckerman.
\newblock A black-box re-weighting analysis can correct flawed simulation data.
\newblock {\em Proceedings of the National Academy of Sciences},
  105(23):7982--7987, 2008.
\newblock {PMCID: PMC2786942}.

\bibitem{Zhang2010c}
C.~Zhang and J.~Ma.
\newblock Enhanced sampling and applications in protein folding in explicit
  solvent.
\newblock {\em The Journal of Chemical Physics}, 132(24):244101, 2010.

\bibitem{Zhang2007b}
J.~Zhang, M.~Lin, R.~Chen, J.~Liang, and J.~S. Liu.
\newblock Monte carlo sampling of near-native structures of proteins with
  applications.
\newblock {\em Proteins}, 66(1):61--68, 2007.

\bibitem{Zhang2008a}
J.~Zhang, M.~Lin, R.~Chen, W.~Wang, and J.~Liang.
\newblock Discrete state model and accurate estimation of loop entropy of rna
  secondary structures.
\newblock {\em The Journal of Chemical Physics}, 128(12):125107, 2008.

\bibitem{Zhang2002}
J.~L. Zhang and J.~S. Liu.
\newblock A new sequential importance sampling method and its application to
  the two-dimensional hydrophobic--hydrophilic model.
\newblock {\em The Journal of Chemical Physics}, 117(7):3492--3498, 2002.

\bibitem{Zhang2010}
X.~Zhang, D.~Bhatt, and D.~M. Zuckerman.
\newblock Automated sampling assessment for molecular simulations using the
  effective sample size.
\newblock {\em Journal of Chemical Theory and Computation}, Accepted., 2010.

\bibitem{Zheng2008a}
L.~Zheng, M.~Chen, and W.~Yang.
\newblock Random walk in orthogonal space to achieve efficient free-energy
  simulation of complex systems.
\newblock {\em Proc Natl Acad Sci U S A}, 105(51):20227--20232, Dec 2008.

\bibitem{Zheng2007a}
W.~Zheng, M.~Andrec, E.~Gallicchio, and R.~M. Levy.
\newblock Simulating replica exchange simulations of protein folding with a
  kinetic network model.
\newblock {\em Proc Natl Acad Sci U S A}, 104(39):15340--15345, Sep 2007.

\bibitem{Zhu2002}
Z.~Zhu, M.~E. Tuckerman, S.~O. Samuelson, and G.~J. Martyna.
\newblock Using novel variable transformations to enhance conformational
  sampling in molecular dynamics.
\newblock {\em Phys Rev Lett}, 88(10):100201, Mar 2002.

\bibitem{Zuckerman2009}
D.~Zuckerman.
\newblock Principles and practicalities of canonical mixed-resolution sampling
  of biomolecules.
\newblock In G.~A. Voth, editor, {\em Coarse-Graining of Condensed Phase and
  Biomolecular Systems}. Taylor \& Francis, Boca Raton, FL, 2009.

\bibitem{Zuckerman-2004d}
D.~M. Zuckerman.
\newblock Simulation of an ensemble of conformational transitions in a
  united-residue model of calmodulin.
\newblock {\em J. Phys. Chem. B}, 108:5127--5137, 2004.

\bibitem{Zuckerman2010}
D.~M. Zuckerman.
\newblock {\em Statistical Physics of Biomolecules: {An} Introduction}.
\newblock CRC Press, 2010.

\bibitem{Zuckerman2006}
D.~M. Zuckerman and E.~Lyman.
\newblock A second look at canonical sampling of biomolecules using replica
  exchange simulation.
\newblock {\em J. Chem. Theory Comput.}, 2(4):1200--1202, 2006.
\newblock {PMCID: PMC2586297}.

\bibitem{Zuckerman-1999}
D.~M. Zuckerman and T.~B. Woolf.
\newblock Dynamic reaction paths and rates through importance-sampled
  stochastic dynamics.
\newblock {\em J. Chem. Phys.}, 111:9475--9484, 1999.

\end{thebibliography}

\end{document}